\documentclass[aps,superscriptaddress,showpacs,amsmath,amssymb,amsfonts,altaffillsymbol,twocolumn,notitlepage,prl]{revtex4-1}

\usepackage{graphicx}
\usepackage{float}
\usepackage{dcolumn}
\usepackage{bm}
\usepackage{color}
\usepackage{soul}
\usepackage{xcolor,hyperref}

\graphicspath{{./figures/}}

\begin{document}

\title{Self-organization and memory in a disordered solid subject to random driving}

\author{Muhittin Mungan}
\email[Corresponding author: ]{mungan@thp.uni-koeln.de}
\affiliation{Institute for Biological Physics, University of Cologne, Z{\"u}lpicher Stra{\ss}e 77, K{\"o}ln, Germany}

\author{D. Kumar}%
\affiliation{PMMH, CNRS, ESPCI Paris, Universit\'e PSL, Sorbonne Universit\'e, Universit\'e Paris Cit\'e, France}%

\author{S. Patinet}%
\affiliation{PMMH, CNRS, ESPCI Paris, Universit\'e PSL, Sorbonne Universit\'e, Universit\'e Paris Cit\'e, France}%

\author{D. Vandembroucq}%
\email[Corresponding author: ]{damien.vandembroucq@espci.fr}
\affiliation{PMMH, CNRS, ESPCI Paris, Universit\'e PSL, Sorbonne Universit\'e, Universit\'e Paris Cit\'e, France}%

\begin{abstract} 
We consider self-organization and memory formation in a mesoscopic model of an amorphous solid subject to a random shear strain protocol confined to a strain range $\pm \varepsilon_{\rm max}$. We develop proper read-out protocols to show that the response of the driven system retains a memory of the strain range, which can be subsequently retrieved. Our findings generalize previous results obtained upon oscillatory driving and suggest that self-organization and memory formation of disordered materials can emerge under more general conditions, such as a disordered system interacting with its fluctuating environment. The self-organization results in a correlation between the dynamics of the system and its environment. We conclude by discussing our results within the context of environmental sensing, highlighting their generalizability to adaptation strategies of simple organisms under changing conditions.
\end{abstract}
\maketitle

\begin{figure}[t!]
\begin{center}
   \includegraphics[width=\columnwidth]{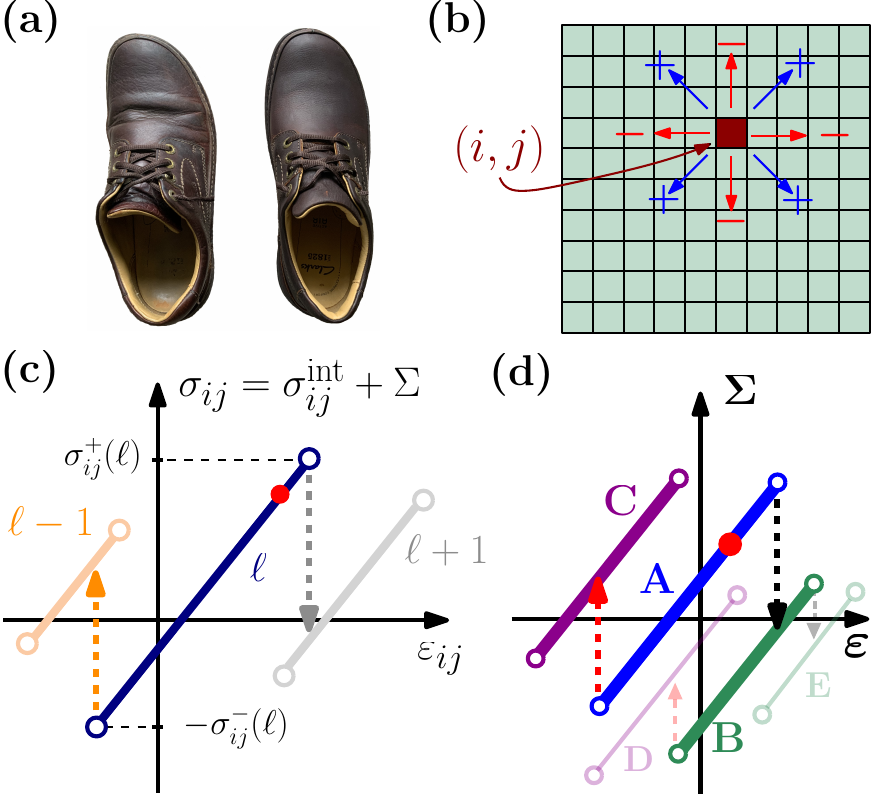} 
    \caption{(a) Right legs from pairs of worn and new shoes of the same make. The worn shoe exhibits a wear and deformation pattern specific to its bearer. (b)
    The yielding of a cell $(i,j)$ of the elastoplastic mesoscale model induces a stress redistribution among the other blocks. (c)  A stack of local elastic branches is associated with a cell $(i,j)$. Each branch $\ell$ is bounded by a pair of local yield stress thresholds $\sigma^\pm_{ij}(\ell)$. 
    Local yield events lead to transitions to neighbouring local elastic branches $\ell \pm 1$. (d) At the macroscopic scale, the system is on a global elastic branch $A$.
    Straining beyond the stability limits $\varepsilon^\pm[A]$, causes a transition to a neighboring branch $B$ or $C$, indicated by dashed arrows. 
    Transitions between such \emph{mesostates} are not necessarily reversible ({\it e.g.} outgoing branches $D$ and $E$ from $B$).  
    }
    \label{fig:epm-model} 
    \end{center}
\end{figure}

Consider two pairs of identical shoes: one has been worn by you for an extended period, while the other has never been worn and is
in pristine condition. As shown in Fig.~\ref{fig:epm-model}(a),
the broken-in shoe will reflect features of you: your gait and, more generally, aspects of your lifestyle, perhaps an active one where you frequently run or jump or a more leisurely one where you walk. Next, imagine lending your worn shoes to a friend who has the same shoe size.
To your friend, the shoes will probably not feel right at first and will require being broken-in once again. As wearers of the shoes, each of you subjects them to mechanical deformations that are unique to you. Consequently, the way you wear the shoes leaves an imprint -- or memory -- on them. Having someone else wear your own shoes will eventually cause some loss of this memory. 

Likewise, disordered materials exhibit a memory of their mechanical past. In experiments on athermal disordered systems driven by oscillatory deformations, such as colloidal
suspensions~\cite{Keimetal2011, Keimetal2014, keim2014mechanical, keim2018return} or crumpled thin elastic sheets~\cite{shohat2022,mungan2022comm,shohat2023dissipation}, 
a self-organized reversible state emerges after several driving cycles. 
Such systems retain a memory of the deformation amplitude, which can subsequently be read-out~\cite{keim2019memory,paulsen2024mechanical}. Self-organization and memory formation under cyclic shear has also been observed in numerical simulations of atomistic as well as mesoscopic models of sheared amorphous solids~\cite{fiocco2014encoding, regev2015reversibility, Priezjev-PRE16, yeh2020glass, bhaumik2021role, chenliu2020, Maloney2021, kumar2022, Liu-Ferrero-JCP22,kumar2024}. 

How can one retrieve the amplitude of past oscillatory driving? An experimentally realizable read-out protocol subjects the ``trained" system to single cycles of oscillatory deformations of increasing amplitude.
By comparing the configuration of the system in the trained state
with the ones obtained at the end of each read-out cycle,
the discrepancy between these two becomes minimal when read-out and training amplitude match~\cite{keim2019memory, paulsen2024mechanical}. 

The example of the broken-in shoe, and more generally of worn clothes, suggests that the mechanical deformation does not have to be strictly oscillatory for a memory of the driving to be imprinted on it. A concept even more pervasive is the fatigue of materials, where repeated stress leads to lasting changes in structure and function~\cite{suresh1998fatigue}.
Here, we present simulations of a 2d elastoplastic model of an amorphous solid subject to a {\em random} shear strain protocol and show that random driving can indeed cause the solid to self-organize into a state that retains memory.

\noindent{\it Elastoplastic Model of a Sheared Amorphous Solid --} We consider the 2d Quenched Mesoscopic Elasto Plastic (QMEP) model of an amorphous solid introduced in~\cite{kumar2022}. Elastoplastic models~\cite{nicolas2018deformation} reproduce the phenomenology of sheared amorphous solids at a quasi-quantitative level~\cite{Patinet-CRP21,Liu-Martens-PRL21,Patinet-ActaMat22,Tyukodi-CRP23}, including the irreversibility transition~\cite{Maloney2021,kumar2022,Liu-Ferrero-JCP22}, and in fact, can also exhibit memory formation and retrieval under oscillatory shear, as 
we have recently shown~\cite{kumar2024}. The solid is coarse-grained into mesoscale blocks, each of which can yield and redistribute local stresses in response to an applied shear strain.
In particular, we associate a stack of local elastic branches with each mesoscale block, labeled by  integers $\ell$, as illustrated in Fig.~\ref{fig:epm-model} (b) and (c). Under loading by athermal quasistatic shear (AQS), the local stress $\sigma_{ij}$ and strain $\varepsilon_{ij}$ of each block follow their branch segment $\ell$ until its termination at $\sigma_{ij}^-(l)$ or $\sigma_{ij}^+(l)$,  depending on the shearing direction. At this point, the mesoscopic block {\em yields}, and a transition to one of the neighbouring branches $\ell \pm 1$ occurs. The local yield event induces a local stress drop and a long-ranged redistribution of internal stresses $\sigma^{\rm int}_{ij}$ among other blocks. The latter follows a quadrupolar pattern, reproducing the effect of an Eshelby inclusion on the surrounding elastic matrix~\cite{eshelby1957determination}. Note that each elastic branch $\ell$ is assigned a given pair of thresholds $\sigma_{ij}^-(\ell), \sigma_{ij}^+(\ell)$, hence the quenched character of the disorder: the same thresholds can be visited several times in a back and forth motion  (refer to \cite{kumar2022} for details on the elastic interaction and the threshold disorder).

At the macroscopic scale, the stress response $\Sigma$ to an externally applied strain $\varepsilon$ consists of a sequence of {\it global} elastic branches punctuated by stress drops. Each elastic branch has a stability range which is determined by the local branch configurations $(\ell_{ij})$ with the requirement that under elastic deformations, each cell $(i,j)$, experiencing the local stress $\sigma_{ij} = \sigma^{\rm int}_{ij} + \Sigma$, 
remains on its local branch~\cite{kumar2022}. We will call these {\it global}  elastic branches {\it mesostates} and label them in capital letters. In Fig.~\ref{fig:epm-model}(d), we illustrate the transitions out of the global elastic branch $A$ to branches $B$ and $C$ upon the increase, respectively decrease, of the applied shear strain. In contrast to local elastic branches, reversibility is not ensured at the macroscopic scale so that after a forward transition $A \to B$, the system can perform a backward transition $B \to D$ and thereby avoid revisiting the global elastic branch $A$.

\begin{figure}[t!]
\begin{center}
\includegraphics[width=\columnwidth]{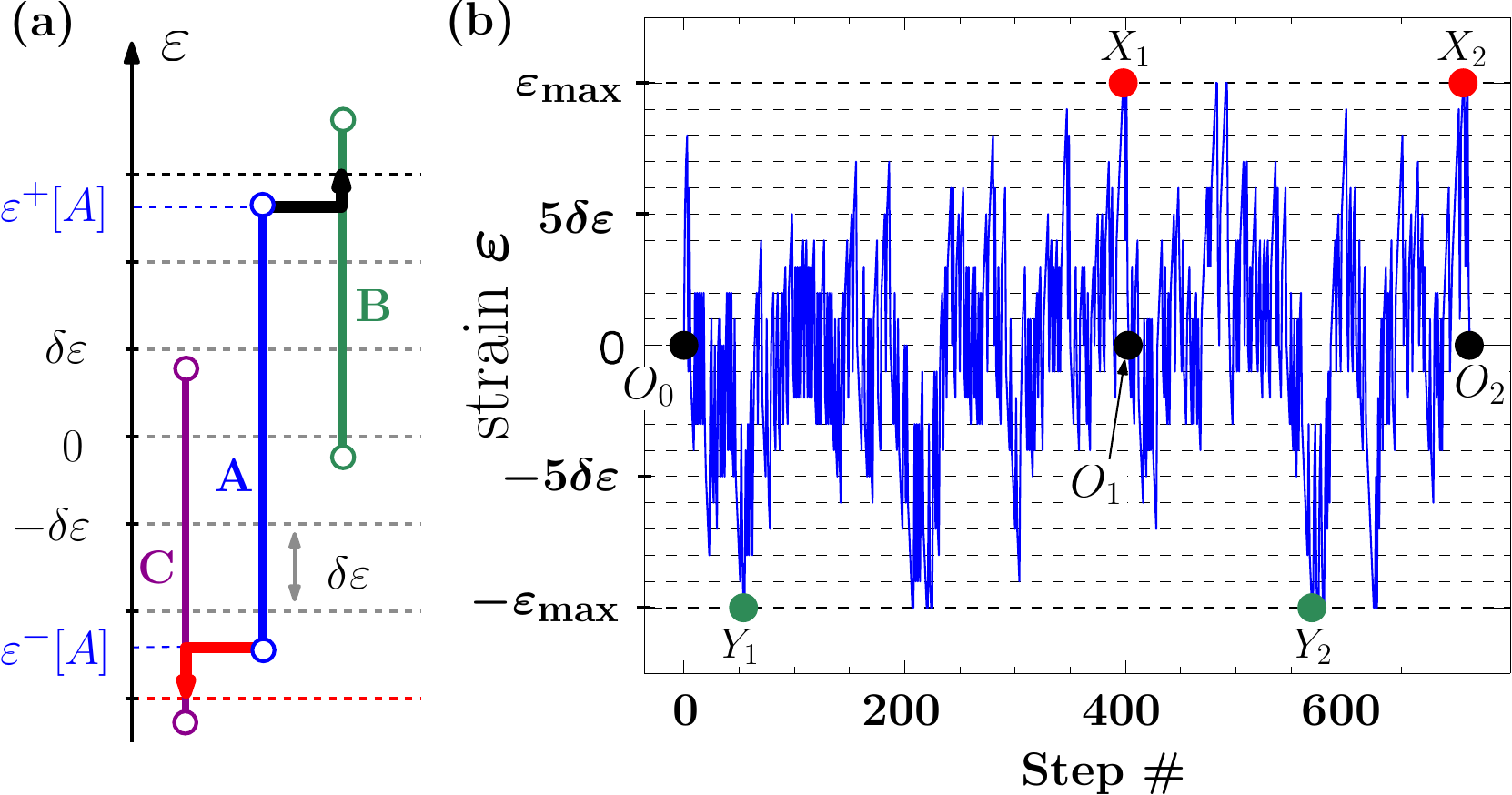}
\caption{(a) The random  
shear driving protocol. The external strain takes fixed values given by multiples of $\delta \epsilon$ and changes randomly. 
The choice of the direction of driving is random and unbiased. (b) A sample random strain driving history shows the applied strain at a given step of the random driving: the system starts in the elastic branch $O_0$ and is subject to random driving with reflecting boundaries at $\pm \varepsilon_{\rm max} = \pm 10 \delta \varepsilon$. The boundary reached first defines the sense of driving and is called the first boundary. A cycle is defined by the first passage times to go from zero strain to the first boundary, then to the opposite boundary and return to zero strain, like the cycle $O_0 \to Y_1 \to X_1 \to O_1$ shown.}
\label{fig:training-combo}
\end{center}
\end{figure}

\noindent{\it Random Driving Protocol --} 
While AQS loading by {\em deterministic shear cycles} $0 \to \varepsilon_{\rm max} \to -\varepsilon_{\rm max} \to 0$ is unambiguously defined, a wide variety of random driving protocols can be considered. Here, we consider a simple random walk along the one-dimensional axis of applied external shear $\varepsilon$, supplied with reflective boundaries at shear strains $\pm \varepsilon_{\rm max}$. Starting at the origin $\varepsilon = 0$, our random driving consists of taking finite strain steps of size $\delta \varepsilon$ with $\varepsilon_{\rm max} = M \delta \varepsilon$, so that it takes at least $M$ steps for the walker to reach one of the reflecting boundaries from the origin. In the following, we deal with strain steps of magnitude $\delta \varepsilon$ greater than the typical stability range $\Delta \varepsilon$ of mesostates. In the case the walker land in a mesostate $A$ whose stability range $\Delta \varepsilon[A] = \varepsilon^+[A] - \varepsilon^-[A]$ is larger than the strain step, that is, when $\Delta \varepsilon [A] > \delta \varepsilon$, we first randomly choose a direction and then apply a strain $k \delta \epsilon$ in that direction, with $k$ such that the resulting strain is closest yet outside the corresponding end of the elastic branch, as illustrated in Fig.~\ref{fig:training-combo}(a). This implementation minimizes the time the walker spends on elastic branches and ensures the symmetry of the random walk
\footnote{The reflecting boundaries at $\pm\varepsilon_{\rm max}$ are implemented as follows: when the walker reaches 
a global elastic branch whose strain range extends beyond the maximal permissible strains, in the next step it reverses its direction, leaving the branch from the boundary that lies within the strain range $[-\varepsilon_{\rm max},\varepsilon_{\rm max}]$.}. 


In analogy to driving by deterministic cyclic shear, we next define 
{\em random} shear cycles. The system starts in a freshly prepared glass $O_0$ at zero applied strain and is then subjected to a random walk along the strain axis, as described already. Fig.~\ref{fig:training-combo}(b) shows a typical deformation path. In this example,  the boundary $- \varepsilon_{\rm max}$ is hit first. We, therefore, define the direction of negative strain as the {\em forward shearing direction} and label the corresponding mesostate when this boundary is hit as $Y_1$. We next track the first time the system reaches the opposite boundary {\em after} $Y_1$ and label the corresponding mesostate $X_1$. The first cycle completes with the first return to zero strain at mesostate $O_1$, as indicated in Fig.~\ref{fig:training-combo}(b). Next, the mesostate $Y_2$ corresponds to the first time the system hits the forward boundary again reaching $O_1$, etc. Our training protocol consists of applying $\mathcal{N}$ random cycles to a fresh glass $O_0$. 
\footnote{Note that the forward passages $O_0 \to Y_1$ or $O_1 \to Y_2$ require far more steps than the reverse ones, i.e.  $X_1 \to O_1$ and $X_2 \to O_2$. This is a direct consequence of the Bauschinger effect~\cite{Patinet-Bauschinger-PRL20}.} 

\noindent{\it Simulation Details --}
We generated $10$ poorly annealed (PA) glasses of size $32\times 32$ mesoscale blocks and subjected them to random driving. Details of the sample preparation protocol can be found in \cite{kumar2022}, where it was shown that for $32\times 32$ PA samples, a cyclic response can be obtained with high probability up to a strain amplitude of $\varepsilon_{\rm irr} \approx 0.05$~\cite{kumar2022}. This value marks the onset of the irreversibility transition~\cite{Pine2005,corte2008random, Keimetal2011, denisov2015sharp, Nagel2017,yeh2020glass, bhaumik2021role, sri2021mesoland, mungan2021metastability, parley2022mean, kumar2022, reichhardt2023reversible}.

\begin{figure}
\begin{center}
\includegraphics[width=0.9\columnwidth]{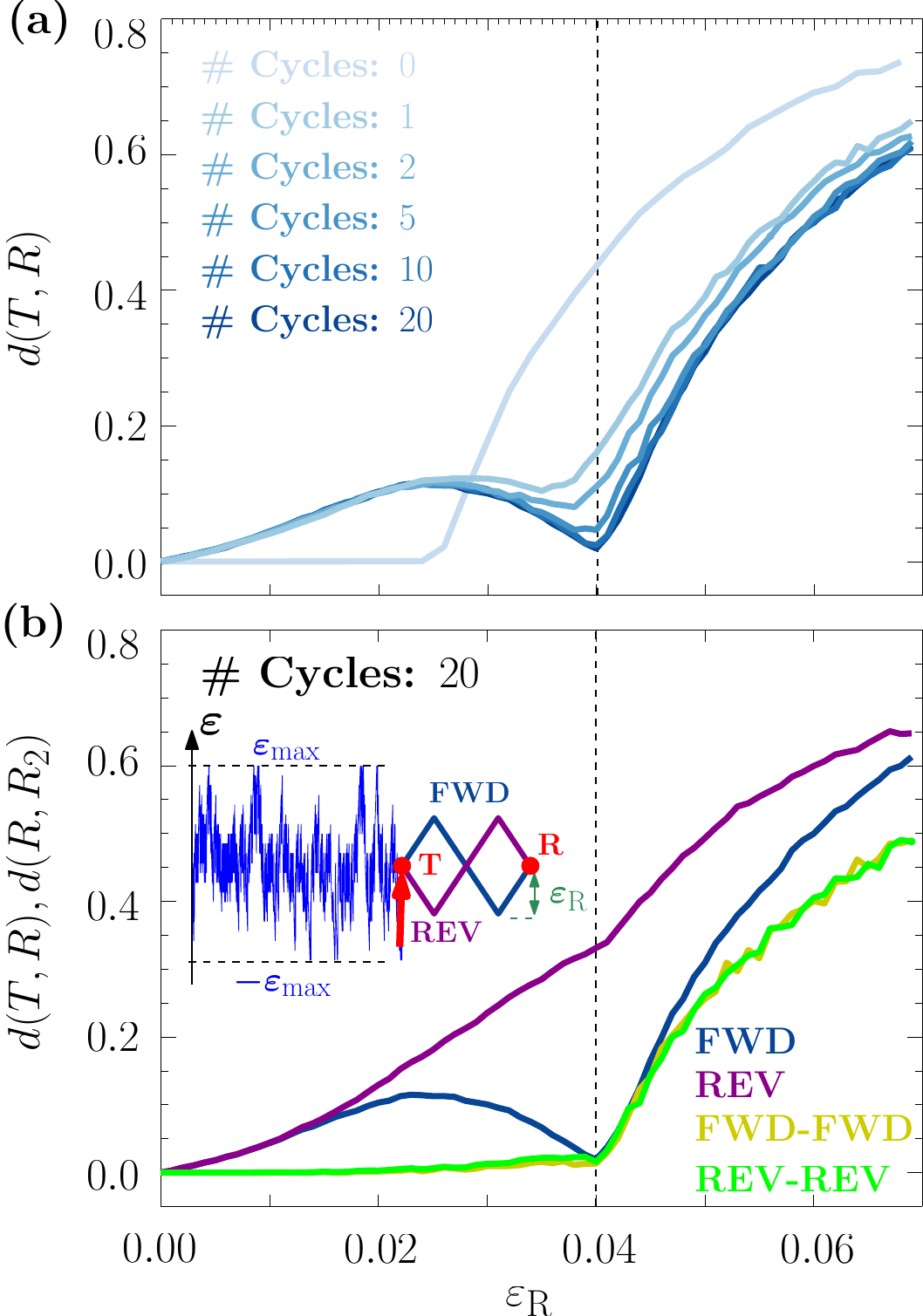}
\end{center}
\caption{(a) Read-out from the trained system after a given number of training cycles. The distance $d(T,R)$ between trained and read-out states shows a cusp at $\varepsilon_{\rm max}  = 0.04$, i.e. where training and read-out amplitudes $\varepsilon_{\rm R}$ coincide. These cusps become more pronounced with the number of training cycles. (b) Behavior of $d(T,R)$ vs. $\varepsilon_{\rm R}$  for four read-out protocols applied to samples subjected to $20$ cycles of random shearing with $\varepsilon_{\rm max} = 0.04$. The inset shows the random training (blue) along with the last direction of approach to the trained state $T$ (red arrow), which determines the FWD/REV sense of the read-out by deterministic oscillatory shear. The FWD-FWD and REV-REV read-outs are obtained by applying two read-out cycles of the same sense to $T$, leading to the state $R_2$ and then tracking $d(R,R_2)$. Except for the out-of-phase protocol REV, all other read-out protocols exhibit a clear change of behavior when $\varepsilon_{\rm R} = \varepsilon_{\rm max}$. }
\label{fig:ro-combo}    
\end{figure}

We considered random driving protocols over strain ranges given by $\varepsilon_{\rm max} = 0.032, 0.036, 0.04, 0.044$ and $0.048$, using random walk step sizes of $\delta \varepsilon = 0.008$, $0.004$, $0.002$ and $0.001$. We drove the system for up to $\mathcal{N} = 250$ random cycles. 
As we describe next, we find that memory features already establish themselves after about 10 driving cycles. In addition, our results on memory encoding and retrieval turn out to be qualitatively similar for all the strain ranges $\varepsilon_{\rm max}$ and step sizes $\delta \varepsilon$ considered. In the following, we therefore show only results for glasses driven for $\mathcal{N} = 20$ cycles with $\varepsilon_{\rm max} = 0.04$ and $\delta \varepsilon = 0.002$. Figs.~\ref{fig:training-various-eps} and \ref{fig:training-various-eps-steps} in the Appendix contain results for the other strain ranges $\varepsilon_{\rm max}$ and steps $\delta \varepsilon$.

\noindent{\it Read-out Protocols --} Having trained our system for a given number $\mathcal{N}$ of driving cycles, we denote the mesostate reached at the end of the driving as the trained state  $T \equiv  O_{\mathcal{N}}$. 
Following previous works on memory retrieval after training by oscillatory shear~\cite{keim2019memory,kumar2024}, we perform a read-out protocol on $T$ by applying a single cycle of {\em deterministic} oscillatory shear at a read-out amplitude $\varepsilon_{\rm R}$, i.e. $0 \to s \varepsilon_{\rm R} \to -s \varepsilon_{\rm R} \to 0$, with $s=\pm 1$ depending on the sense of the read-out. 

Suppose first that we know the {\em sense} of the random cycles 
established during training, i.e.  whether the $+\varepsilon_{\rm max}$ or $-\varepsilon_{\rm max}$ boundary was reached first, or equivalently, hit last. We can have a read-out that is {\em in-phase} with the sense of training~\cite{kumar2024}, which we will call the forward (FWD) read-out, so that the initial shear direction points
{\em towards} the boundary hit first. Conversely, if the direction of read-out strain initially points {\em opposite} to it, and hence it is {\em out-of-phase} with the training, this will be referred to as a reverse (REV) read-out. For the example shown in Fig.~\ref{fig:training-combo}(b), where the $-\varepsilon_{\rm max}$ boundary was reached first, $0 \to -\varepsilon_{\rm R} \to \varepsilon_{\rm R} \to 0$ correspond to a FWD read-out cycle, while $0 \to \varepsilon_{\rm R} \to -\varepsilon_{\rm R} \to 0$ is a REV read-out. Regardless of the sense of direction, we refer to the state reached at the end of the first read-out cycle as $R$. 

Next, we define a distance between the two mesostates $T$ and $R$ by comparing their corresponding sets of local branch indices $\ell_{ij}[T]$ and $\ell_{ij}[R]$ as follows: 
\begin{equation}
    d(T,R) = \frac{1}{N^2} \, \# \left \{ (ij) : \vert  \ell_{ij}[T] - \ell_{ij}[R] \vert > 0 \right \},
    \label{eqn:ddef}
\end{equation}
i.e. we determine the fraction of cells that fail to return to their local elastic branches at the end of the read-out cycle.

\noindent{\it Memory after Random Driving --}  Fig~\ref{fig:ro-combo}(a) shows the behavior of $d(T,R)$ with FWD read-out at various amplitudes $\varepsilon_{\rm R}$ and for different numbers $\mathcal{N}$ of driving cycles for glasses trained at $\varepsilon_{\rm max}= 0.04$. The data points are averages over $10$ realizations, and we consider a {\em parallel} read-out protocol~\cite{fiocco2014encoding,kumar2024}: for each glass, we keep a copy of $T$ and then apply the read-outs at different amplitudes to the same  $T$. An experimentally realizable {\em sequential} read-out protocol leads to similar results (see Appendix Figs.~\ref{fig:sequential-ro-a} and \ref{fig:sequential-ro-b}). 

As a control, we also performed a read-out on the freshly prepared glasses $O_0$, i.e. a state not yet subjected to any shear. The distance $d(T,R)$ remains zero up to a read-out strain of $\varepsilon_{\rm R} = 0.024$. This value is consistent with the typical stability range $\Delta \varepsilon[T]$ of the un-trained state, indicating that over this range of read-outs our glass will likely respond purely elastically and thus return to the initial configuration after one read-out cycle. The remaining graphs in Fig~\ref{fig:ro-combo}(a) show the results of read-outs after $\mathcal{N} = 1, 2, 5, 10$ and $20$ random driving cycles. Already for $\mathcal{N} = 1$, a local minimum of $d(T,R)$ appears near the training amplitude $\varepsilon_{\rm max} = 0.04$ \footnote{Note that similar memory signatures were observed already after two driving cycles in MD simulations under deterministic oscillatory shear~\cite{fiocco2014encoding}. In addition, a random driving cycle typically triggers a larger number of transitions between elastic branches than a deterministic cycle. For the maximum strain $\varepsilon_{\rm max} = 0.04$ shown, the random and deterministic driving cycles comprise, on average, $4960$ and $400$ transitions, respectively. Consequently, there is more mechanical annealing during a random driving cycle.}. With increasing $\mathcal{N}$, this dip becomes cusp-like and moves towards $\varepsilon_{\rm max}$. Note that for $\mathcal{N} = 10$ and $20$, the corresponding curves are nearly indistinguishable. We have verified that this behavior does not change appreciably when the random driving is continued to $\mathcal{N} = 250$ cycles. 

We performed the read-out in the FWD direction to recover the training amplitude $\varepsilon_{\rm max}$. As mentioned above, this requires knowing the sense of the driving established 
during training. We next ask how $d(T,R)$ behaves if we use the reverse read-out protocol REV instead. This is shown in Fig~\ref{fig:ro-combo}(b), where we compare the results of the FWD (dark blue) and REV (magenta) read-out protocols applied to $T = O_{20}$. We see that the REV protocol is not able to detect well $\varepsilon_{\rm max}$, if at all \footnote{In the case of deterministic oscillatory shear and for training amplitudes well below the irreversibility transition, a memory of the shear amplitude can be detected in REV read-outs, as revealed by a sudden increase of the slope of $d(T,R)$~\cite{kumar2024}. Similarly, in Fig.~\ref{fig:training-various-eps}(a) for $\varepsilon_{\rm max} = 0.032$ and $0.036$, a kink of $d(T,R)$ at $\varepsilon_{\rm R} \approx \varepsilon_{\rm max}$ is clearly discernible, but gets fainter with increasing $\varepsilon_{\rm max}$. Note that the FWD and REV read-outs observed in Figs.~\ref{fig:ro-combo}(b) and Fig.~\ref{fig:training-various-eps}(a) are consistent with what one would obtain if return-point memory held, as discussed in detail in \cite{kumar2024} }. This finding is consistent with our recent work on memory formation in the QMEP model under deterministic oscillatory shear~\cite{kumar2024}. 

\noindent{\it Two-Cycle Read-out Protocols --}  The read-out procedure introduced so far requires prior knowledge of the sense of driving. 
In the following, we take advantage of the emergent reversibility of the trained samples to propose a two-cycle read-out protocol that does not require such information. 
The sequence of plastic events during training induces a statistical hardening so that, within the range of stresses explored by the system, the material gets increasingly harder and  therefore exhibits more and more  pseudo-elastic behavior~\cite{BVR-PRL02,TPVR-Meso12,Patinet-Bauschinger-PRL20,kumar2022}. This evolution comes with the development of a quasi {\em return-point-memory} (RPM) behavior
\footnote{RPM is a property of magnets with ferromagnetic spin interactions and elastic manifolds driven through random disorder, where the stress distribution upon triggering an instability is systematically destabilizing~\cite{middleton1992asymptotic}. Owing to the anisotropic nature of stress distribution in sheared amorphous solids, localized yielding events can stabilize or destabilize other parts of the sample. Hence, RPM is not expected to hold a priori in such systems. Nevertheless, recent work on driven disordered systems has demonstrated experimentally as well as numerically that these systems can exhibit near-perfect RPM \cite{keim2018return, mungan2019networks, shohat2022,shohat2023dissipation, kumar2024}. }, and characterizes 
the ability of a system to return to a state at which the direction of driving has been reverted~\cite{sethna1993hysteresis, munganterzi2018,terzi2020state}.

To exploit this emergent reversibility behavior, we define a two-cycle read-out protocol as follows. Denote by $R_2$ the mesostate reached when applying to $R$ an additional deterministic read-out cycle at the same amplitude and sense of direction as the first. If RPM were to hold, then $d(R,R_2) = 0$, {\em for any} amplitude $\varepsilon_R$ and state $T$ to which the two-cycle read-out is applied, i.e. regardless of whether $T$ is a result of some prior training or not~\cite{middleton1992asymptotic, sethna1993hysteresis, munganterzi2018} (see also note \footnote{More precisely, $d(R,R_2) \equiv 0$ over the full admissible range of driving amplitudes. In the case of ferromagnets, the range of admissible field amplitudes has no limit, as the magnetization will saturate at sufficiently large fields. However, in the case of elastic manifolds driven through random media  AQS conditions do not hold anymore when these manifolds depin, limiting the range of driving under which $d(R,R_2) \equiv 0$.
}). Thus, instead of considering $d(T,R)$, we observe the behavior of $d(R,R_2)$ with read-out amplitude. Depending on the sense chosen, we will refer to these as the FWD-FWD and REV-REV read-out protocols. The ochre (green) curve in Fig~\ref{fig:ro-combo}(b) shows the read-outs under the FWD-FWD (REV-REV) protocols. Note that both protocols are insensitive to the sense of driving, exhibiting near indistinguishable behavior of $d(R,R_2)$ with $\varepsilon_{\rm R}$. Moreover,  $d(R,R_2) \approx 0$ up to the maximum training strain $\varepsilon_{\rm max} = 0.04$, after which it rises sharply. 

Thus, when restricted to $\varepsilon_{\rm R} \lesssim \varepsilon_{\rm max}$, we find that the trained glasses exhibit a highly mechanically reversible response, while for read-out strains larger than $\varepsilon_{\rm max}$, reversibility is rapidly lost.
In the absence of any training direction information, applying the two-cycle reading protocol allows us to retrieve the maximum amplitude of the material's past random driving.

Whether and how this mechanical reversibility is due to an emergent RPM property restricted to the range of prior training is a point left for future work.

\noindent{\it Discussion --} 
Our findings show that memory formation and retrieval in amorphous solids can be achieved under random driving, 
thereby generalizing previous results obtained under oscillatory shear~\cite{Keimetal2011, fiocco2014encoding, keim2019memory, paulsen2024mechanical, kumar2024}.  
Thus it is not the driving protocol itself but the range of applied strains, as captured by $\varepsilon_{\rm max}$, that encodes the memory.  

The random driving of the amorphous solid causes a self-organization into a state of high mechanical reversibility. 
This is most clearly seen by comparing the distances after single-cycle read-outs FWD and REV  
with those after two-cycle read-outs, FWD-FWD and REV-REV, Figs.~\ref{fig:ro-combo}(b) and Fig.~\ref{fig:training-various-eps}(b):  the second read-out cycle always returns the system to the vicinity of the state $R$ reached at the end of the first read-out cycle {\em regardless} of how far $R$ is away from the trained state $T$. This behavior persists as long as the applied strains $\vert \varepsilon \vert \lesssim \varepsilon_{\rm max}$. The emergent mechanical reversibility is, therefore, marginal. 
 
More generally, we can ask about the nature of emergent self-organization, the possibility of dynamical attractors into which the random driving may steer and trap the system, as well as  the nature of reversibility underlying the memory formation. These are questions that are most conveniently addressed via the transition graph ($t$-graph) description of the AQS response of driven systems~\cite{mungan2019networks, Regev2021, kumar2022}. This will be left for future work.

We considered a simple random walk with finite strain steps and reflecting boundaries at well-defined strains $\pm \varepsilon_{\rm max}$. The nature of the real fluctuations experienced by materials in working conditions is clearly more complex. 
For example, one could relax the boundary condition and test the effect of spatial and temporal correlations in random driving or add thermal noise to simulate aging. In the present case, energy is injected at a large scale, while   active matter can be  regarded as random driving operating at a small scale~\cite{agoritsas2024memory, goswami2023yielding, alert2022activeturbulence}. It would be instructive to compare the processes of self-organization and memory formation   in these contrasting conditions. Probing the adaptive response of active biological tissues to random mechanical actuation is of interest as well~\cite{laplaud2021pinching}. Other natural perspectives of this work are learning in disordered networks~\cite{stern_physical_2024} and the fatigue behavior of engineering materials under service~\cite{morrow_cyclic_1965}. 

The everyday life example of worn shoes naturally invites experimental testing of mechanical self-organization and memory under random driving. Experimentally implementable read-out protocols are available, and as we have shown for the amorphous solid, these lead to qualitatively similar results, cf. Figs.~\ref{fig:sequential-ro-a} and \ref{fig:sequential-ro-b}.
Keeping with the theme of breaking-in shoes, an application of these ideas to  textiles, such as the fading of jeans by natural (or artificial) wear, might be of interest as well.

We conclude by taking a step back to discuss possible broader implications of our findings. A natural setting of random driving is that of a system interacting with its fluctuating environment, and our results imply that the latter can leave an imprint on the former.

The random driving leads to a steady-state in which the subsequent response of the system is correlated with the driving history. As we have seen, this correlation persists as long as the magnitude of loading does not exceed that encountered during the training. The resulting mechanical reversibility implies that the sequence of mechanical instabilities triggered, and their locations in the sample are persistently entangled with random driving. 
As a result, state changes within the sample encode some information about the dynamic changes in its environment. 
In other words, the training by random driving established an internal representation of the system's environment to which it is coupled: the dynamics of the environment evolution is mirrored by the pattern of instabilities triggered by the former \cite{libchaber2020walking}. We may therefore think of the self-organization  under random driving  as having led to a capability that in the context of machine learning would be associated with pattern recognition, unsupervised learning, and more generally, environmental sensing.

It is generally believed that even the adaptation of relatively simple organisms, such as bacteria, involves some form of information processing, memory and learning~\cite{Murugan_2021}. Given that a model of a disordered system with minimal ingredients -- disorder, a built-in frustration in redistribution of external loads and negligible thermal effects -- is able to self-organize in a manner described above, one may wonder whether simple biological organisms lacking a nervous system, exploit this essentially ``for free" infrastructure when adapting to changing environments. Put differently~\cite{batten2008visions}, can {\em self-organization propose what natural selection disposes}?

\begin{acknowledgements}
The authors would like to thank Srikanth Sastry, Yoav Lahini, Dor Shohat, Hugo Fort and Elisabeth Agoritsas for useful discussions. MM also acknowledges PMMH and ESPCI for their kind hospitality during multiple stays, which were made possible by Chaire Joliot grants awarded to him. MM was supported by the Deutsche Forschungsgemeinschaft (DFG, German Research Foundation) under Projektnummer 398962893 and also partly under CRC 1310 {\em Predictability in Evolution}. This project has received funding from the European Union’s
Horizon 2020 research and innovation programme under the Marie
Sklodowska-Curie grant agreement No. 754387.
\end{acknowledgements}

\bibliographystyle{apsrev4-1}
\bibliography{AmorphNets_extnd}

\newpage

\appendix
\section*{Appendix}
\renewcommand{\thefigure}{A\arabic{figure}}
\setcounter{figure}{0}

\subsection*{Parallel read-outs at different $\varepsilon_{\rm max}$}

\begin{figure}[b!]
\begin{center}
   \includegraphics[width=0.85\columnwidth]{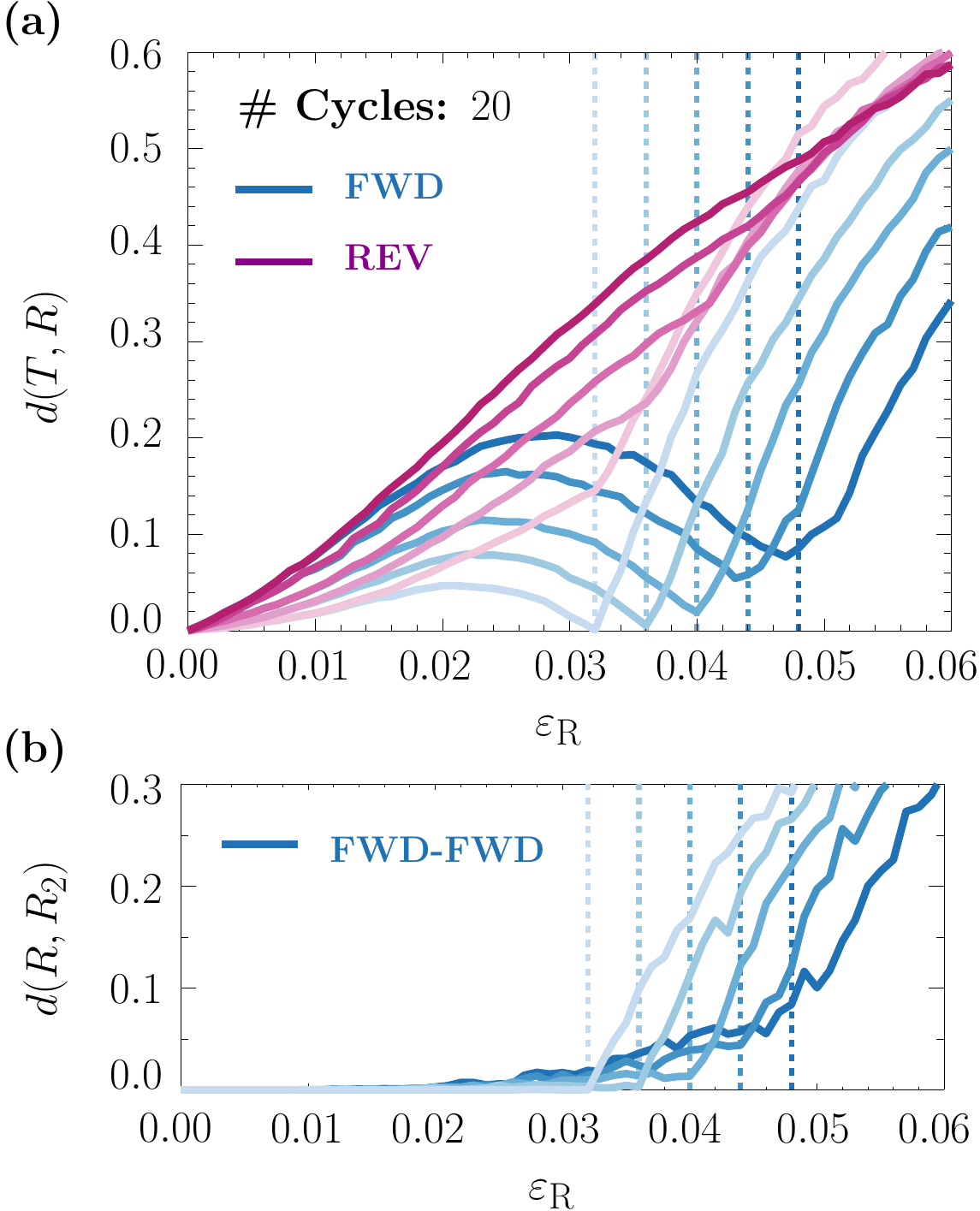} \\
    \caption{Read-out from states $T$ reached after $20$ random driving cycles. 
    (a) The result of FWD (blue tones) and REV (red tones) outs, with the darkness of the color increasing with $\varepsilon_{\rm max}$. (b) Read-outs using the FWD-FWD protocol. 
    }
    \label{fig:training-various-eps} 
    \end{center}
\end{figure}

Keeping the step size at $\delta \varepsilon = 0.002$, we subjected  $10$ poorly-aged glasses to random driving confined by  maximal strains: $\varepsilon_{\rm max} = 0.032, 0.036, 0.04, 0.044$ and $0.048$. Fig.~\ref{fig:training-various-eps}(a) and (b) show the results of performing parallel read-outs of type FWD, REV and FWD-FWD. Shades of darker colors denote larger values of $\varepsilon_{\rm max}$. Vertical dashed lines indicate the latter.

The FWD read-outs in panel (a) show a local minimum of the read-out distance $d(T,R)$ at amplitudes $\varepsilon_{\rm R}$ close to the maximum strain $\varepsilon_{\rm max}$ of the random driving. With increasing $\varepsilon_{\rm R}$, the distance $d_{\rm min}$ at the local minimum starts to depart from a value close to zero. The distances $d(T,R)$ of the REV protocol increase monotonically with $\varepsilon_{\rm R}$, however, for $\varepsilon_{\rm max} \lesssim 0.04$, they show a discernible kink at $\varepsilon_{\rm R} \approx \varepsilon_{\rm max}$, see also~\cite{kumar2024}.

Panel (b) of Fig.~\ref{fig:training-various-eps} shows the evolution of $d$ under the FWD-FWD read-out protocol. For maximum strain values up to $\varepsilon_{\rm max} = 0.04$, the distances $d$ remain small and close to zero. This behavior continues up to $\varepsilon_{\rm R} \approx \varepsilon_{\rm max}$, beyond which $d(R,R_2)$ starts to increase abruptly, giving rise to a kink in the read-out curves. 
The REV-REV read-outs show statistically similar behavior.

\subsection*{Random driving with different step sizes $\delta \varepsilon$}

\begin{figure}[b!]
\begin{center}
   \includegraphics[width=0.85\columnwidth]{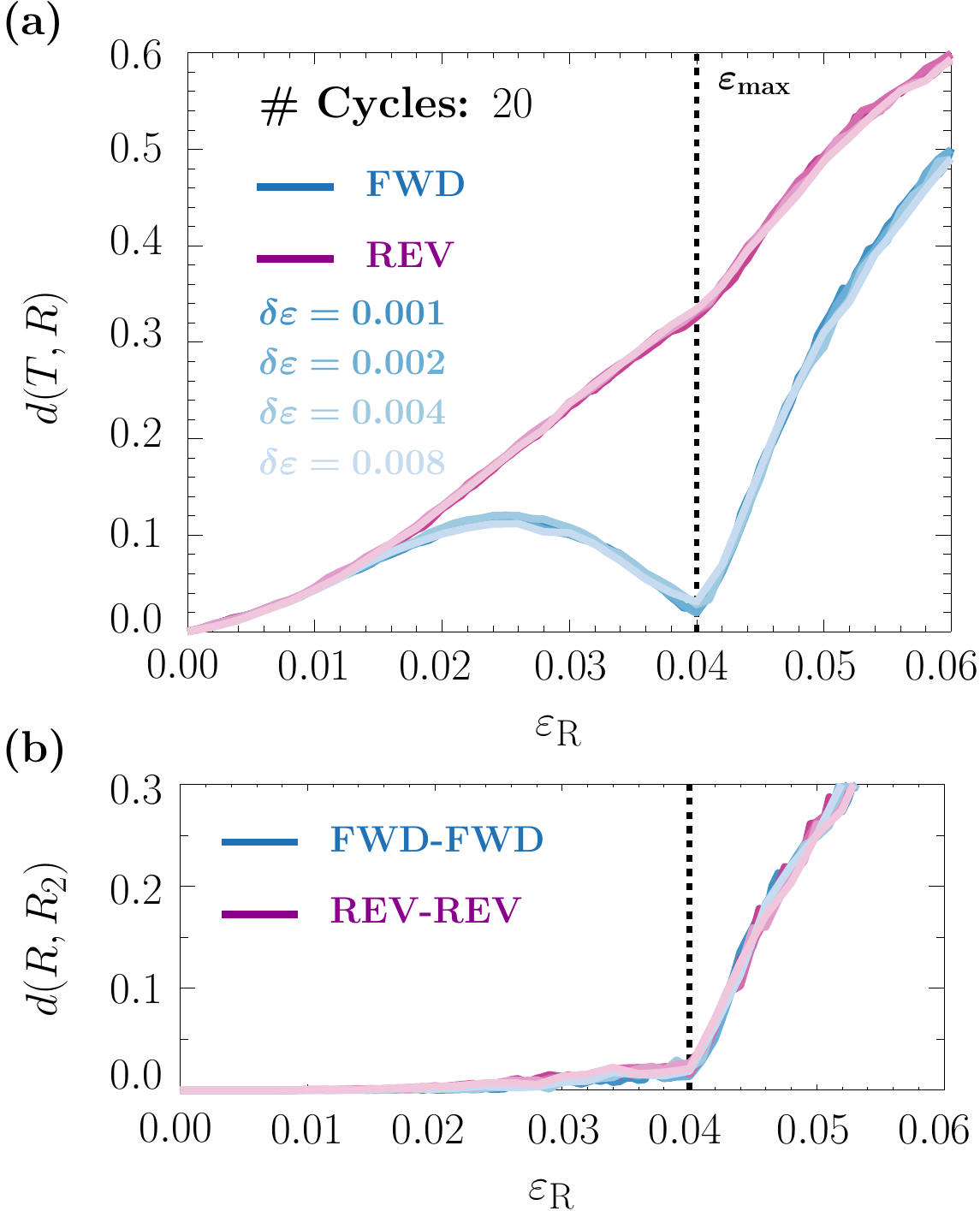} \\
    \caption{Read-out from states $T$ reached after $20$ random driving cycles with $\varepsilon_{\rm max} = 0.04$ fixed, but using different random walk strain step sizes $ \delta \varepsilon = 0.008, 0.004, 0.002$ and $0.001$, color coded in shades from light to dark. 
    (a) The result of FWD (blue tones) and REV (red tones) outs, with the darkness of the color increasing with $\varepsilon_{\rm max}$. (b) Read-outs using the FWD-FWD (blue tones) and REV-REV protocol (red tones). 
    }
    \label{fig:training-various-eps-steps} 
    \end{center}
\end{figure}

We have also subjected our 10 poorly-aged glasses to random driving protocols where we kept the strain range constant at $\varepsilon_{\rm max} = 0.04$, but used different values for the random walk strain step size: $ \delta \varepsilon = 0.008, 0.004, 0.002$ and $0.001$. Fig.~\ref{fig:training-various-eps-steps} shows the results of performing read-outs after 20 random cycles for: (a) FWD and REV read-outs, color coded in tones of blue and red, respectively, with the shading getting lighter with increasing random walk step size, and (b), FWD-FWD and REV-REV read-outs, using the same color coding as in (a). The dependence on step size is minimal.

\begin{figure}
\begin{center}
   \includegraphics[width=\columnwidth]{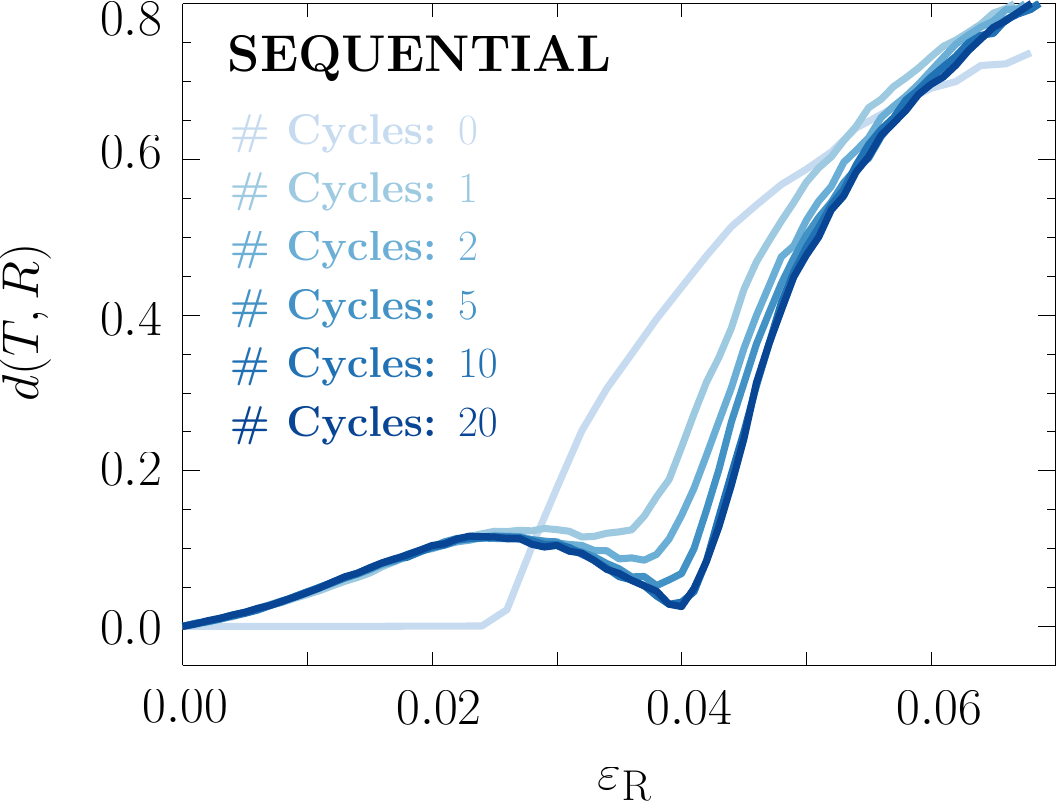} \\
   \caption{Read-out from the trained states $T$ using a sequential read-out protocol. The trained glasses are the same as the ones used for the read-outs in Figs.~\ref{fig:ro-combo}(a) and (b). Shown is the evolution of the read-out quality with the number of training cycles.  The results are qualitatively similar to those obtained from parallel read-outs in Fig.~\ref{fig:ro-combo}(a). 
    }
   \label{fig:sequential-ro-a} 
   \end{center}
\end{figure}

\begin{figure}
\begin{center}
   \includegraphics[width=\columnwidth]{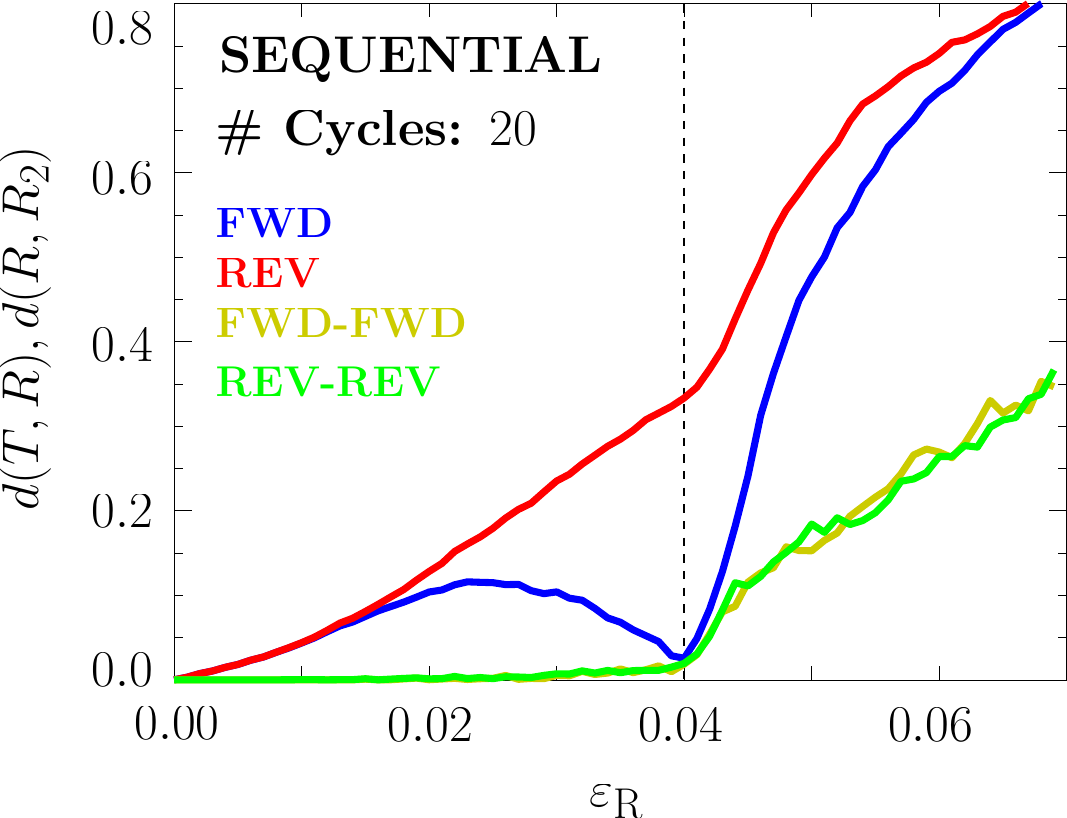} \\
   \caption{Read-out from the trained states $T$ using a sequential read-out protocol. The trained glasses are the same as the ones used for the read-outs in Figs.~\ref{fig:ro-combo}(a) and (b). The results of read-out using the four different protocols. The results are qualitatively similar to those obtained from parallel read-outs in Fig.~\ref{fig:ro-combo}(b). 
    }
   \label{fig:sequential-ro-b} 
   \end{center}
\end{figure}

\subsection*{Random driving with sequential read-outs}
The read-out protocols defined in the main text are parallel: it is assumed that we can restore the trained state $T$ for each read-out amplitude. Experimentally, this is not easy to achieve. An alternative is to perform {\em sequential read-outs}~\cite{kumar2024}. We start with the trained state and consider a sequence of increasing read-out amplitudes $\varepsilon_{\rm R}^{(n)}$. Specifically, we start our read-out with $\varepsilon_{\rm R}^{(1)}$ applied to $T$: we apply two cycles of cyclic shear and record the configurations $R^{(1)}$ and $R^{(1)}_2$ at the end of the first and second read-out cycle. We then increase the read-out amplitude to $\varepsilon_{\rm R}^{(2)}$ and apply the two-cycle read-out to $R^{(1)}_2$ and continue in this manner through our list of amplitudes. As shown in Figs.~\ref{fig:sequential-ro-a} and \ref{fig:sequential-ro-b}, this type of read-out gives qualitatively similar results to the parallel case that was shown in Figs.~\ref{fig:ro-combo} (a) and (b).

\end{document}